\documentclass[10pt,superscriptaddress,twocolumn,amsmath,amssymb,aps,prl]{revtex4}
\usepackage{mathrsfs}
\usepackage{graphicx}
\usepackage{dcolumn}
\usepackage{bm}

\newcommand{\be}{\begin{equation}}
\newcommand{\ee}{\end{equation}}
\newcommand{\bea}{\begin{eqnarray}}
\newcommand{\eea}{\end{eqnarray}}

\begin{document}
\title{Evolution of Higgs mode in a Fermion Superfluid with Tunable Interactions}
\author{Boyang Liu}
\affiliation{Institute for Advanced Study, Tsinghua University,
Beijing, 100084, China}

\author{Hui Zhai}
\affiliation{Institute for Advanced Study, Tsinghua University,
Beijing, 100084, China}

\author{Shizhong Zhang}
\affiliation{Department of Physics and Center of Theoretical and
Computational Physics, The University of Hong Kong, Hong Kong,
China}

\date{\today}
\begin{abstract}
In this letter we present a coherent picture for the evolution of Higgs mode in both neutral and charged $s$-wave fermion superfluids, as the strength of attractive interaction between fermions increases from the BCS to the BEC regime. In the case of neutral fermionic superfluid, such as ultracold fermions, the Higgs mode is pushed to higher energy while at the same time, gradually loses its spectral weight as interaction strength increases toward the BEC regime, because the system is further tuned away from Lorentz invariance. On the other hand, when damping is taken into account, Higgs mode is significantly broadened due to coupling to phase mode in the whole BEC-BCS crossover. In the charged case of electron superconductor, the Anderson-Higgs mechanism gaps out the phase mode and suppresses the coupling between the Higgs and the phase modes, and consequently, stabilizes the Higgs mode.
\end{abstract}
 \maketitle

The experimental search for Higgs boson in particle physics has made remarkable progresses~\cite{Cern1, Cern2}. On the other hand, Higgs mode has also generated considerable interest in condensed matter and cold atom systems. Early in 1980s', Raman scattering experiment has revealed an unexpected peak in a superconducting charge density wave compound $\text{NbSe}_2$~\cite{Sooryakumar}, which was later attributed to the Higgs mode~\cite{Littlewood1, Littlewood2}. Signal of Higgs mode has also been observed in antiferromagnet $\text{TlCuCl}_3$ by the neutron scattering~\cite{Ruegg}, and recently in superconducting NbN sample by terahertz pump probe spectroscopy in a nonadiabatic excitation regime~\cite{Matsunaga1, Matsunaga2}. In cold atom system, Higgs mode has been observed near the superfluid to Mott insulator phase transition of bosonic atoms in optical lattices at integer filling~\cite{Bissbort, Endres}.

Theoretically, the simplest field theory where Higgs mode emerges is a relativistic $U(1)$ field theory with Lorentz invariance in the symmetry broken phase. This occurs, for example, in the weak coupling BCS superconductor~\cite{Varma2002,Barlas2013} or in the Mott-superfluid transition of Bose-Hubbard model at integer filling~\cite{Huber2008,Pollet2012}.  However, in most condensed matter systems, Lorentz invariance only emerges with fine tuning and the generic symmetry is usually Galilean~\cite{Pekker2014}. Thus, it is an interesting question to investigate how the Higgs mode evolves as the system is tuned away from the Lorentz invariance point. Moreover, in condensed matter systems, further complications often occur because the Higgs mode is usually coupled to other elementary excitations which leads to its damping~\cite{Podolsky2011,Podolsky2012,Gazit2013,Rancom2014}. In this Letter, we investigate these issues in the context of the BEC-BCS crossover model. In the BCS limit, the system obeys approximate Lorentz symmetry due to particle-hole symmetry and is expected to host Higgs mode. In the BEC limit, it is a condensate of molecular bosons and obeys the Galilean invariance. It thus provides a unique system to describe the fate of Higgs mode as the system is tuned away from Lorentz invariant limit. In addition, due to tunable interactions in the BEC-BCS crossover, it also provides a great platform to investigate the interaction effects on the Higgs mode due to coupling to collective and quasi-particle excitations~\cite{Bruun2014}.

\begin{figure}[t]
  \includegraphics[width=0.4\textwidth]{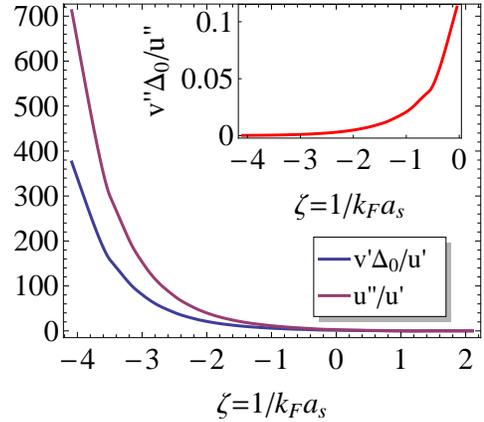}
  \caption{(Color online) $v'\Delta_0/u'$ and $u''/u'$ as functions of the scattering length $\zeta=1/k_F a_s$.
  In the inset we show $v''\Delta_0/u''$ as a function of $\zeta$. }
 \label{fig:uv}
\end{figure}

We investigate these questions based on the time-dependent
Ginzburg-Landau formulation of the BEC-BCS crossover, \be S=\int dt
d^3{\bf
x}\big[\phi^*(-iu\partial_t+v\partial_t^2-\frac{\nabla^2}{2m^\ast}-r)\phi+\frac{b}{2}|\phi|^4\big],
\label{eq:GL} \ee where $\phi$ is the Ginzburg-Landau order
parameter. The various parameters $u,v,r,b$ and $m^*$ can be
computed along BEC-BCS crossover in terms of the chemical potential
$\mu$, temperature $T$ and $\zeta=1/(k_{\rm F}a_{\rm S})$, where $a_{\rm S}$ is the s-wave scattering length. Within
the Nozi\`{e}res-Schmit-Rink \cite{NSR} framework, this can be
calculated as detailed in the supplementary material~\cite{supp}.
The coefficients of the time derivative terms $u=u'+iu''$ and
$v=v'+iv''$ are complex in general. The real parts $u'$ and $v'$
describe the propagating behavior of the cooper pair field, while
the imaginary parts $u''$ and $v''$ describe its damping due to
coupling to the fermionic quasi-particles. A plot of various
parameters are given in Fig.\ref{fig:uv}. We note the following
features.

(i) Consider the real parts $u'$ and $v'$ in the BEC-BCS crossover. In the BCS limit, $u'/v'\Delta_0\to 0$ because of the approximate particle-hole symmetry in the weak-coupling BCS theory while $\Delta_0=\sqrt{r/b}$ is the mean field value of order parameter. As a result, the system acquires an emergent Lorentz invariance, and one expects the emergence of Higgs mode, together with the standard Anderson-Bogoliubov mode for neutral fermion superfluid. In the BEC limit, however, $v'\Delta_0/u'\sim \Delta_0/|\mu|\ll 1$, and we can neglect the $v'$-term. This leads to a Galilean invariant neutral boson theory, for which only Bogoliubov mode exists.

(ii) The damping terms ($u''$) becomes important as one moves to the BCS side, because of the decreasing fermionic excitation gap and as a result, a stronger coupling of the pairing field to the quasi-particle excitations. This corresponds to finite lifetime of Cooper pairs at finite temperature. We will show that the damping $u''$-term itself will generate considerable effect for the appearance of the Higgs mode different from that in a pure Lorentz invariance theory. In the BEC limit, the imaginary parts vanishes within NSR. On the other hand, we find that whenever they are nonzero, $v''\Delta_0/u''\ll 1$ for the entire crossover regime and we shall thus neglect $v''$-term altogether in the following discussion.

\begin{figure}[t]
   \includegraphics[width=0.48\textwidth]{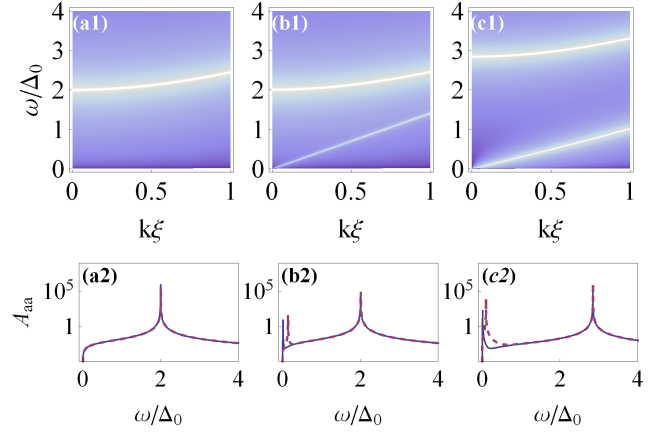}
  \caption{(Color online) Spectral function $A_{aa}({\bf k},\omega)$ in the absence of damping term. $A_{aa}({\bf k},\omega)$ as a function of $k$ (in unit of $1/\xi$) and $\omega$
  (in unit of $\Delta_0$) for three different interaction strength $\zeta=-1/(k_Fa_s)$, $\zeta=-7$ for (a), $\zeta=-3$ for (b) and $\zeta=-1$ for (c), corresponding to
  different gaps
   $\Delta_0/E_F=10^{-5}$, $\Delta_0/E_F=4\times10^{-3}$ and $\Delta_0/E_F=7\times10^{-2}$, respectively.
   (a2-c2): $A_{aa}({\bf k},\omega)$ as a function of $\omega$ for $k=0.1/\xi$ (purple dashed line) and $k=0.01/\xi$ (blue solid line). $T/T_c=0.9$ and $\delta$ is taken as $10^{-4}\Delta_0$. }
 \label{fig:spectralweight}
\end{figure}

{\em Spectral Weight Transfer without Damping.} To investigate the evolution of Higgs mode as the system is tuned gradually from its Lorentz-invariant BCS limit towards the Galilean invariant BEC limit, we shall first neglect the damping terms in Eq.\ref{eq:GL} and study the transfer of spectral weight between the Higgs and Goldstone modes. In the symmetry broken state, we can write the order parameter $\phi=\Delta_0+\delta_a+i\delta_p$, where $\delta_a$ and $\delta_p$ describe amplitude and phase fluctuations, respectively. In terms of $\delta_a$ and $\delta_p$ and with $u''=v''=0$, we can write the action Eq. \ref{eq:GL} in the Fourier space as
\be
S=\int\frac{d\omega}{2\pi}\frac{d^3{\bf k}}{(2\pi)^3}\bar{\Phi}(-\omega,-{\bf k})\mathcal{G}^{-1}\Phi(\omega,{\bf k}),
\ee
with $\bar{\Phi}(\omega,{\bf k})=(\delta_a(\omega,{\bf k}),\delta_p(\omega,{\bf k}))$ and the kernel $\mathcal{G}$ is given by
\begin{equation}
\mathcal{G}^{-1}=\left(
\begin{array}{cc}
-v'\omega^2+\xi_{\bf k}+2r & iu'\omega\\
-iu'\omega & -v'\omega^2+\xi_{\bf k}
\end{array}\right),
\end{equation}
with $k=|{\bf k}|$ and $\xi_{\bf k}=k^2/2m^*$. Two branches of spectrum can be identified, with mode frequencies given by
\be
\omega_\pm^2=\frac{\xi_{\bf k}+r}{v'}+\frac{u'^2}{2v'^2}\pm\sqrt{\frac{r^2}{v'^2}+\frac{u'^4}{4v'^4}+\frac{u'^2}{v'^3}(\xi_{\bf k}+r)}.
\ee

In the BCS limit, $v'\Delta_0\gg u'$ and solutions can be written as $\omega_-(k)=k/\sqrt{2m^*v'}$ and $\omega_+(k)=\sqrt{(\xi_{\bf k}+2r)/v'}$; the first being the Goldstone mode with linear dispersion and the second Higgs mode, with Higgs gap $\omega_+(0)=\sqrt{2r/v'}=\sqrt{2b\Delta_0^2/v'}$. Using the facts that $v'=7\beta^2\zeta(3)\nu_0/16\pi^2$ and $b=7\beta^2\zeta(3)\nu_0/8\pi^2$ in the BCS limit, one finds $\omega_+(0)=2\Delta_0$, as expected for a Lorentz-invariant theory. Here $\beta=1/k_BT$ is the inverse temperature, $\zeta(n)$ is the Riemann-Zeta function and $\nu_0$ is the density of state at the Fermi energy $\epsilon_F$.

In the BEC limit, $u'\gg v'\Delta_0$, we find $\omega_-(k)=\sqrt{\xi_{\bf k}(\xi_{\bf k}+2r)}/u'$ is the Bogoliubov mode while the other mode $\omega_+(k)=\sqrt{2\xi_{\bf k}/v'+2r/v'+(u'/v')^2}$ has a gap $\sim |\mu|$, of order of binding energy of the molecule in the BEC limit. The existence of the gapped mode is a reflection of the fact that our bosonic field $\phi$ is a composite of two fermions and disappears in the infinite binding limit where only Bogoliubov mode exists as it should.

In between these two limits, Lorentz invariance is broken and the
coupling between the amplitude and phase degrees of freedom becomes
stronger, as characterized by the off-diagonal term $iu'\omega$. We
note that for low energy Bogoliubov excitations, such coupling is
small, but for gapped Higgs mode, it provides significant mixing of
the amplitude and phase. To characterize such mixing, we calculate
the spectral function for the amplitude $\delta_a$, given by
$A_{aa}({\bf k},\omega)=-\frac{1}{\pi}{\rm Im}\mathcal{G}_{aa}({\bf
k},\omega+i\delta)$. Explicitly, this can be written as \be
A_{aa}({\bf k},\omega)=A_+({\bf k})\delta(\omega-\omega_{+}({\bf
k}))+A_-({\bf k})\delta(\omega-\omega_{-}({\bf k}))
\label{spectralnd} \ee with $A_+({\bf k})$ and $A_-({\bf k})$ being
the spectral weight densities associated with two modes $\omega_-$
and $\omega_+$~\cite{supp}. In Fig.\ref{fig:spectralweight} (a,b,c), we plot the
spectral function $A_{aa}({\bf k},\omega)$ for three representative
values of $\zeta$ (corresponding to different
$\Delta_0/E_\text{F}$). Two features can be noticed immediately.
First, the Higgs gap increases beyond $2\Delta_0$ of the BCS limit as interaction strength increases. Secondly,
there is increasing spectral weight transfer from the gapped Higgs
mode to the gapless mode. One can show explicitly that
$A_{+}/A_{-}=4v'r^2(u'^2\sqrt{k^2/2m^\ast(k^2/2m^\ast+2r)})^{-1}$,
which indicates the gradual increasing of the mixing between phase
and amplitude degrees of freedom.

{\em Including Damping Term.} Due to the presence of damping term, the time-dependent Ginzburg-Landau theory is not a pure Lorentz invariant $U(1)$ theory. Thus, at any finite temperature, even in the BCS limit, the peak of Higgs excitation $\omega_+({\bf k})$ will not as sharp as discussed above. To calculate the equilibrium spectral weight in the presence of damping, we need to introduce the so-called Langevin force $\eta(t,{\bf x})$, which satisfies the following conditions, $\langle\eta(t',{\bf x}')\eta(t,{\bf x})\rangle=\langle\eta^*(t',{\bf x}')\eta^*(t,{\bf x})\rangle=0$,
and $\langle\eta^*(t',{\bf x}')\eta(t,{\bf x})\rangle=2u''k_B T\delta(t-t')\delta({\bf x}-{\bf x}')$. Including the corresponding term in the action as $S_L=\int dt d^3{\bf x}(\phi^*\eta+\phi \eta^*)$, we obtain the equations of motion for $\delta_a$ and $\delta_p$, by setting $\partial(S+S_L)/\partial \delta_a=0$ and $\partial(S+S_L)/\partial \delta_p=0$,
\begin{align}
\left(-v\omega^2+\xi_{\bf k}+2r\right)\delta_a-iu\omega\delta_p+\eta' &=0,\\
\left(-v\omega^2+\xi_{\bf k}\right)\delta_p+iu\omega\delta_a+\eta'' &=0,
\end{align}
where $\eta'$ and $\eta''$ are the real and imaginary parts of the Langevin force $\eta$, respectively. The spectral functions for the  amplitude fluctuation is given by, using fluctuation dissipation theorem,
\begin{align}
A_{aa}=\frac{u''\omega}{2}\frac{|-v\omega^2+\xi_{\bf k}|^2+|u\omega|^2}{|-(u\omega)^2+(-v\omega^2+\xi_{\bf k})(-v\omega^2+\xi_{\bf k}+2r)|^2}.
\end{align}

\begin{figure}[t]
 \includegraphics[width=0.48\textwidth]{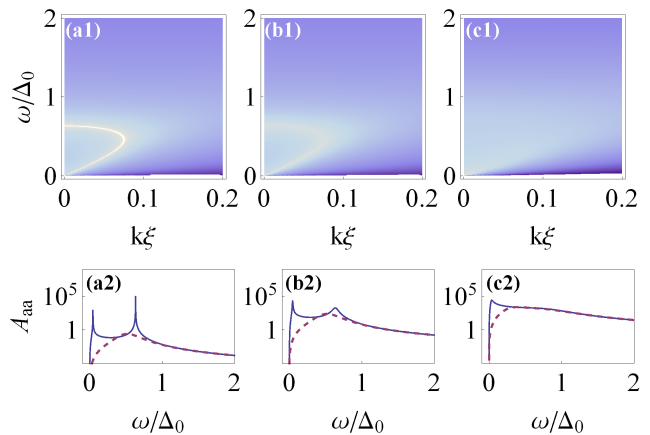}
  \caption{(Color online) Spectral function $A_{aa}({\bf k},\omega)$ in presence of damping term. $A_{aa}({\bf k},\omega)$ as a
   function of $k$ (in unit of $1/\xi$) and $\omega$ (in unit of $\Delta_0$) for three different interaction strength $\zeta=-1/(k_Fa_s)$, $\zeta=-7$ for
    (a), $\zeta=-3$ for (b) and $\zeta=-1$ for (c), corresponding to different  $\Delta_0/E_F=10^{-5}$, $\Delta_0/E_F=4\times10^{-3}$ and $\Delta_0/E_F=7\times10^{-2}$, respectively.
   (a2-c2): $A_{aa}({\bf k},\omega)$ as a function of $\omega$ for $k=0.1/\xi$ (purple dashed line) and $k=0.01/\xi$ (blue solid line). $T/T_c=0.2$ and $\delta$ is taken as $10^{-4}\Delta_0$. }
\label{fig:spectral} \end{figure}

By comparing Fig. \ref{fig:spectral} with Fig. \ref{fig:spectralweight}, one can see three important features brought about by including the damping term. First, the spectral weight transfer is enhanced. For instance, for $\zeta=-7$, there is almost no spectral weight transfer in the absence of damping (Fig. \ref{fig:spectralweight}(a)) while in the presence of damping, for very small $k\ll 1/\xi$, $A_{aa}({\bf k},\omega)$ exhibits a clear peak at the energy of Bogoliubov mode, with a weight proportional to $u''$ \cite{supp}. Similar enhancement of spectral weight transfer can also be easily seen in Fig. \ref{fig:spectral}(b) for $\zeta=-3$. Secondly, also for $k\ll 1/\xi$, in the BCS limit, the location of Higgs peak is substantially reduced from $\sqrt{2r/v'}$ to $\sqrt{2r/v'-u''^2/v'^2}$, as shown for $\zeta=-7$ and $-3$ in Fig. \ref{fig:spectral}(a) and (b), respectively~\cite{supp}.  Thirdly, as $k$ starts to derivate from zero, the Higgs mode quickly loses its identity, due to strong hybridization with the Bogoliubov mode. For instance, even for $k=0.1/\xi$, as displayed by the purple dashed line in Fig. \ref{fig:spectral}(a2-c2), no feature of sharp peak is observed in $A_{aa}({\bf k},\omega)$. And for $\zeta=-1$, no sharp peak exists even for $k=0.01/\xi$.

\begin{figure}[t]
 \includegraphics[width=0.48\textwidth]{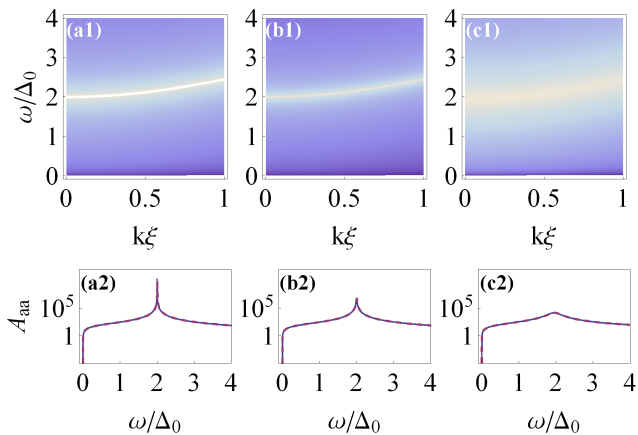}
  \caption{(Color online) Spectral function $A_{aa}({\bf k},\omega)$ for the charged case.
   $A_{aa}({\bf k},\omega)$ as a function of $k$ (in unit of $1/\xi$) and $\omega$ (in unit of $\Delta_0$)
    for three different interaction strength $\zeta=-1/(k_Fa_s)$, $\zeta=-7$ for
    (a), $\zeta=-3$ for (b) and $\zeta=-1$ for (c), corresponding to different  $\Delta_0/E_F=10^{-5}$, $\Delta_0/E_F=4\times10^{-3}$ and $\Delta_0/E_F=7\times10^{-2}$, respectively.
   (a2-c2): $A_{aa}({\bf k},\omega)$ as a function of $\omega$ for $k=0.1/\xi$ (purple dashed line) and $k=0.01/\xi$ (blue solid line). $T/T_c=0.9$ and $\delta$ is taken as $10^{-4}\Delta_0$.}\label{fig:spetralcoul}
\end{figure}

{\em Effects of Coupling to External Gauge Fields}. Now we understand that, in the weakly interacting BCS side of a neutral superfluid, the appearance of Higgs mode suffers significant broadening due to finite $u''$-term at finite temperature, which couples the Higgs mode to the collective Bogoliubov excitations. Therefore, if we further consider the presence of coupling to external electromagnetic field for the case of charged fermions, the Bogoliubov mode is gapped out by the Anderson-Higgs mechanism. Thus, we expect that the Higgs mode is easier to observe in the charged case. To incorporate this effect of external electromagnetic field, we introduce the gauge potential $\varphi(t,{\bf x})$ and extend the action as
\begin{align}
S_{\rm c}&=\int dt d^3{\bf x}\big\{\phi^*[-iu(\partial_t-2e\varphi)+v(\partial_t-2e\varphi)^2\\\nonumber
&-\frac{\nabla^2}{2m^\ast}-r]\phi+\frac{b}{2}|\phi|^4-\frac{1}{8\pi}\varphi\nabla^2\varphi\big\},
\end{align}
where $e$ is the charge of the electron. Following the same procedure as before, we find that the coupling between $\delta_a$ and $\delta_p$ is modified and is now proportional to $k^2$
\begin{align}
i\frac{2u\omega k^2}{k^2+32\pi ve^2\Delta_0^2}\delta_a(\omega,{\bf k})\delta_p(-\omega,-{\bf k}).
\end{align}
As a result, the original gapless phase mode is gapped to a finite frequency, $\omega({\bf k})=\sqrt{\xi_{\bf k}/v'+16\pi e^2\Delta^2_0/m^*}$, which is known as the Anderson-Higgs mechanism. Thus, the large energy separation between this gapped phase mode and Higgs mode strongly suppresses their coupling.  A further consequence of the modification is that at long wave length $k\to 0$, the coupling between phase and amplitude mode becomes small. The spectral function $A_{aa}({\bf k},\omega)$ for charged case is plotted in Fig. \ref{fig:spetralcoul}~\cite{supp}. In sharp contrast to neutral case Fig. \ref{fig:spectral}, the presence of damping term has almost no effect on Higgs mode, and there is always a peak located at $\omega=2\Delta_0$. In this case, as attractive interaction increases and the system gradually loses its Lorentz invariance, the peak becomes more and more broad.

{\em Conclusion.} In summary, we have investigated the evolution of Higgs mode in the BEC-BCS crossover for both neutral and charged Fermi superfluid. Our main conclusions include: i) Towards the BEC side, as the system gradually loses the Lorentz invariance, the Higgs mode is pushed to very high energy and the spectral weight is transferred to Bogoliubov mode. ii) In the BCS side, damping terms arises in the Ginzburg-Landau theory, due to coupling between Cooper pair field and the fermionic quasi-particles, and strongly couples the Higgs mode to the gapless phase mode in the neutral superfluid, which enhances the spectral weight transfer and washes out features of Higgs mode at finite momentum. (iii) For the charged case, the phase mode is gapped out by coupling to external electromagnetic field, and the Higgs mode becomes much more stable.

Our results also deepen our understandings of Higgs mode in superconductor. The physical picture behind the observation of Higgs mode in a BCS superconductor is much more subtle and its observability is not merely guaranteed by Lorentz symmetry. While the damping terms broadens the Higgs peak, the Anderson-Higgs mechanism alleviate the coupling between Higgs and phase mode and as a result, Higgs mode remains at energy $2\Delta_0$.  As for cold atom system, because of the cooling limit, so far we can not reach Fermi superfluid for $\zeta<-1$. However, our results show no Higgs feature in spectral function for $\zeta>-1$. On the other hand, with recent development of synthetic gauge field, there are many proposals to generate a synthetic dynamic gauge field in cold atom system~\cite{Goldman2014}. If such a dynamic gauge field can be experimentally realized and coupled to fermions, the Anderson-Higgs mechanism will be activated and a Higgs mode will be observed. This can be used as a way to test our theory.

\emph{Acknowledgements.} BY and HZ are supported by Tsinghua
University Initiative Scientific Research Program, NSFC Grant No. 11174176,
No. 11325418 and NKBRSFC under Grant No. 2011CB921500. SZ is supported
by a start-up grant from University of Hong Kong, the Collaborative Research Fund
 HKUST3/CRF/13G and RGC-GRF 17306414. HZ would like to thank Hong Kong
 University for hospitality where this work was initiated.

\pagebreak

\begin{widetext}
\setcounter{equation}{0} \setcounter{figure}{0}
\setcounter{table}{0} \setcounter{page}{1} \makeatletter

\section{Supplementary Materials}
\subsection{Time-dependent Ginzburg-Landau theory of BEC-BCS
crossover}

A time-dependent Ginzburg-Landau theory can be constructed for the
entire BEC-BCS crossover in the vicinity of $T_{\rm
c}$~\cite{Sademelosupp}. The partition function takes the form
$\mathcal Z=\int
D[\bar\psi_\sigma,\psi_\sigma]e^{-S[\bar\psi_\sigma,\psi_\sigma]}$,
with \be S[\bar\psi_\sigma,\psi_\sigma]=\int d\tau d^3{\bf x}\Big
\{\bar\psi_\sigma(\partial_\tau-\frac{\nabla^2}{2m}-\mu)\psi_\sigma-g\bar\psi_\uparrow
\bar\psi_\downarrow\psi_\downarrow\psi_\uparrow\Big \}, \ee where
$\psi_\sigma$ are Grassman fields and $g$ is the contact interaction
between fermions of opposite spins. $\mu$ is the chemical potential
which is determined by requiring the number density to be equal to
$n$. To investigate the fluctuation effects in the Cooper channel,
we use a Hubbard-Stratonovich transformation to decouple the
interaction term in the Cooper channel and then integrating out the
fermions. We obtain an effective theory for the bosonic field
$\Delta(\tau, {\bf x})$, which represents the cooper pair field.
Straightforward calculations yield the partition function in terms
of field $\Delta$ as \be \mathcal Z=\int D(\bar\Delta,\Delta)\exp
\Big[-\frac{1}{g}\int d\tau d{\bf x}|\Delta|^2+\ln\det\hat
G^{-1}\Big], \ee where \be \hat{G}^{-1}=
\left(\begin{array}{cc}-\partial_\tau+\frac{\nabla^2}{2m}+\mu &
\Delta\\ \bar \Delta &
-\partial_\tau-\frac{\nabla^2}{2m}-\mu\end{array}\right) \ee
 is the Gor'kov Green function.

In the vicinity of the phase transition the gap parameter $\Delta$
is small and an expansion in terms of $\Delta$ becomes possible.
Including both the spatial and time derivatives (after Wick
rotation) and retaining the parameter $\Delta$ up to the forth order
we obtain an effective action as \be S[\bar\Delta,\Delta]=\int dt
d^3{\bf
x}\Big\{\bar\Delta\big[-iu\partial_t+v\partial_t^2-\frac{\nabla^2}{2m^\ast}-r\big]\Delta+\frac{b}{2}\bar\Delta
\bar\Delta\Delta\Delta\Big\}, \label{eq:GL} \ee where $u=u'+iu''$
and $v=v'+iv''$ are complex in general and all the parameters can be
expressed in terms of microscopic parameters as
\begin{align}
u'
&=\frac{(2m)^{3/2}}{16\pi^2}\Bigg[\frac{2\sqrt2\beta\sqrt{|\mu|}}{\pi}\sum_{n=0}^\infty\frac{\sqrt{\sqrt{1+(2n+1)^2(\frac{\pi}{\beta\mu})^2}-\mbox{sgn}(\mu)}}{(2n+1)^2}
-\frac{\pi\beta}{2}\sqrt{|\mu|}\theta(-\mu)\Bigg],\\
u'' &=\frac{m^{3/2}}{8\sqrt2\pi}\beta\sqrt {|\mu|}\Theta(\mu),\\
v'&=\frac{(2m)^{3/2}}{32\pi^2}\Bigg[\frac{2\sqrt2\beta^2\sqrt{|\mu|}}{\pi^2}\sum_{n=0}^\infty\frac{\sqrt{\sqrt{1+(2n+1)^2(\frac{\pi}{\beta\mu})^2}+\mbox{sgn}(\mu)}}{(2n+1)^3}
-\frac{\pi\beta}{4\sqrt{|\mu|}}\theta(-\mu)\Bigg],\\
v''&=-\frac{m^{3/2}}{32\sqrt2\pi}\frac{\beta}{\sqrt {|\mu|}}\Theta(\mu),\\
\frac{1}{2m^\ast} &=\frac{1}{2m}\int\frac{d^3{\bf
k}}{(2\pi)^3}\Bigg\{\frac{1-2N(\xi_{{\bf k}})}{8\xi_{{\bf k}}^2}
+\frac{\frac{\partial N(\xi_{{\bf k}})}{\partial\xi_{\bf k}}}{4\xi_{\bf k}}+\frac{\frac{\partial^2N(\xi_{{\bf k}})}{\partial \xi^2_{\bf k}}\cdot\frac{{\bf k}^2}{2m}}{6\xi_{\bf k}}\Bigg\},\\
 r &=\frac{m}{4\pi a}+\int \frac{d^3{\bf k}}{(2\pi)^3}\Bigg\{\frac{1-2N(\xi_{\bf k})}{2\xi_{\bf k}}-\frac{1}{2\epsilon_{\bf k}}\Bigg\},\\
b &=\int\frac{d^3{\bf k}}{(2\pi)^3}\Bigg\{\frac{1-2N(\xi_{\bf
k})}{4\xi_{\bf k}^3}+\frac{\beta N(\xi_{\bf k})[N(\xi_{\bf
k})-1]}{2\xi_{\bf k}^2}\Bigg\}. \label{eq:GLparameter}
\end{align}
In the above equations, $N(\xi_{\bf k})=1/(\exp(\beta\xi_{\bf
k})+1)$ is the Fermi distribution function and $\xi_{\bf
k}=\epsilon_{\bf k}-\mu$ with $\epsilon_{\bf k}={\bf k}^2/2m$.
Function $\Theta(2\mu)$ is the heaviside step function. Explicitly,
the parameter $b$ is the result of one-loop calculation with four
fermion propagators \be
b=-\frac{1}{\beta^2}\sum_{\omega_n}\int\frac{d^3{\bf
k}}{(2\pi)^3}\frac{1}{(-i\omega_n+k^2/2m-\mu)^2}\frac{1}{(i\omega_n+k^2/2m-\mu)^2}.
\ee The other parameters $u$, $v$, $\frac{1}{2m^\ast}$ and $r$ are
all derived from the inverse vertex function
$\Gamma^{-1}(\omega_n,{\bf k})$, which after the standard
renormalization by replacing $g$ with the two-body scattering length
$a_{\rm s}$, is given by \be \Gamma^{-1}(\omega_n,{\bf
k})=-\frac{m}{4\pi a_{\rm s}}-\int \frac{d^3{\bf
k}}{(2\pi)^3}\left\{\frac{1-N(\epsilon_{\bf k}-\mu)-N(\epsilon_{\bf
k-q}-\mu)}{-i\omega_n+\epsilon_{\bf k}+\epsilon_{\bf
k-q}-2\mu}-\frac{1}{2\epsilon_{\bf k}}\right\}. \ee To derive the
time-dependent Ginzburg-Landau equation, we first analytically
continue vertex function to real frequency
$i\omega_n\to\omega+i0^+$. This procedure generates a time-dependent
term with parameter $u$ and $v$. The detailed derivation is as
following.

The frequency dependent part of $\Gamma^{-1}(\omega,{\bf k})$ is
\bea \Gamma^{-1}(\omega, 0)-\Gamma^{-1}(0,0)=-\frac{m}{4\pi a}-\int
\frac{d^3{\bf k}}{(2\pi)^3}\Bigg\{\frac{1-2N(\epsilon_{\bf
k}-\mu)}{-\omega-i\eta+2\epsilon_{\bf
k}-2\mu}-\frac{1}{2\epsilon_{\bf k}}\Bigg\}-\Gamma^{-1}(0,0).\eea
Then we expand it in series of small $\omega$ as \bea
\Gamma^{-1}(\omega, 0)-\Gamma^{-1}(0,0)\simeq-\omega\cdot\int
\frac{d^3{\bf k}}{(2\pi)^3}\frac{1-2N(\epsilon_{\bf
k}-\mu)}{(2\epsilon_{\bf k}-2\mu-i\eta)^2}-\omega^2\cdot\int
\frac{d^3{\bf k}}{(2\pi)^3}\frac{1-2N(\epsilon_{\bf
k}-\mu)}{(2\epsilon_{\bf k}-2\mu-i\eta)^3}.\eea We define the
parameters as $u\equiv\int \frac{d^3{\bf
k}}{(2\pi)^3}\frac{1-2N(\epsilon_{\bf k}-\mu)}{(2\epsilon_{\bf
k}-2\mu-i\eta)^2}$ and $v\equiv\int \frac{d^3{\bf
k}}{(2\pi)^3}\frac{1-2N(\epsilon_{\bf k}-\mu)}{(2\epsilon_{\bf
k}-2\mu-i\eta)^3}$. They both can be calculated by contour
integration. \bea u&&\equiv\int\frac{d^3{\bf
k}}{(2\pi)^3}\frac{1-2N(\epsilon_{\bf k}-\mu)}{(2\epsilon_{\bf
k}-2\mu-i\eta)^2}\cr&&=\frac{\nu(\epsilon_F)}{4\sqrt{\epsilon_F}}\int^\infty_0
d\epsilon\sqrt{\epsilon}\frac{1-2N(\epsilon-\mu)}{(\epsilon-\mu-i\eta)^2}\cr&&=\frac{\nu(\epsilon_F)}{4\sqrt{\epsilon_F}}\cdot\frac{1}{2}\int_c
dz\sqrt{z}\frac{1-2N(z-\mu)}{(z-\mu-i\eta)^2},\label{contour}\eea
where $c$ denotes the contour in Fig. \ref{fig:contour}. There are
infinite first-order poles $z_n=\mu+\frac{(2n+1)\pi i}{\beta}$ and
one second order pole $z_\eta=\mu+\eta i$.
\begin{figure}[h]
\begin{center}
  \includegraphics[width=5cm]{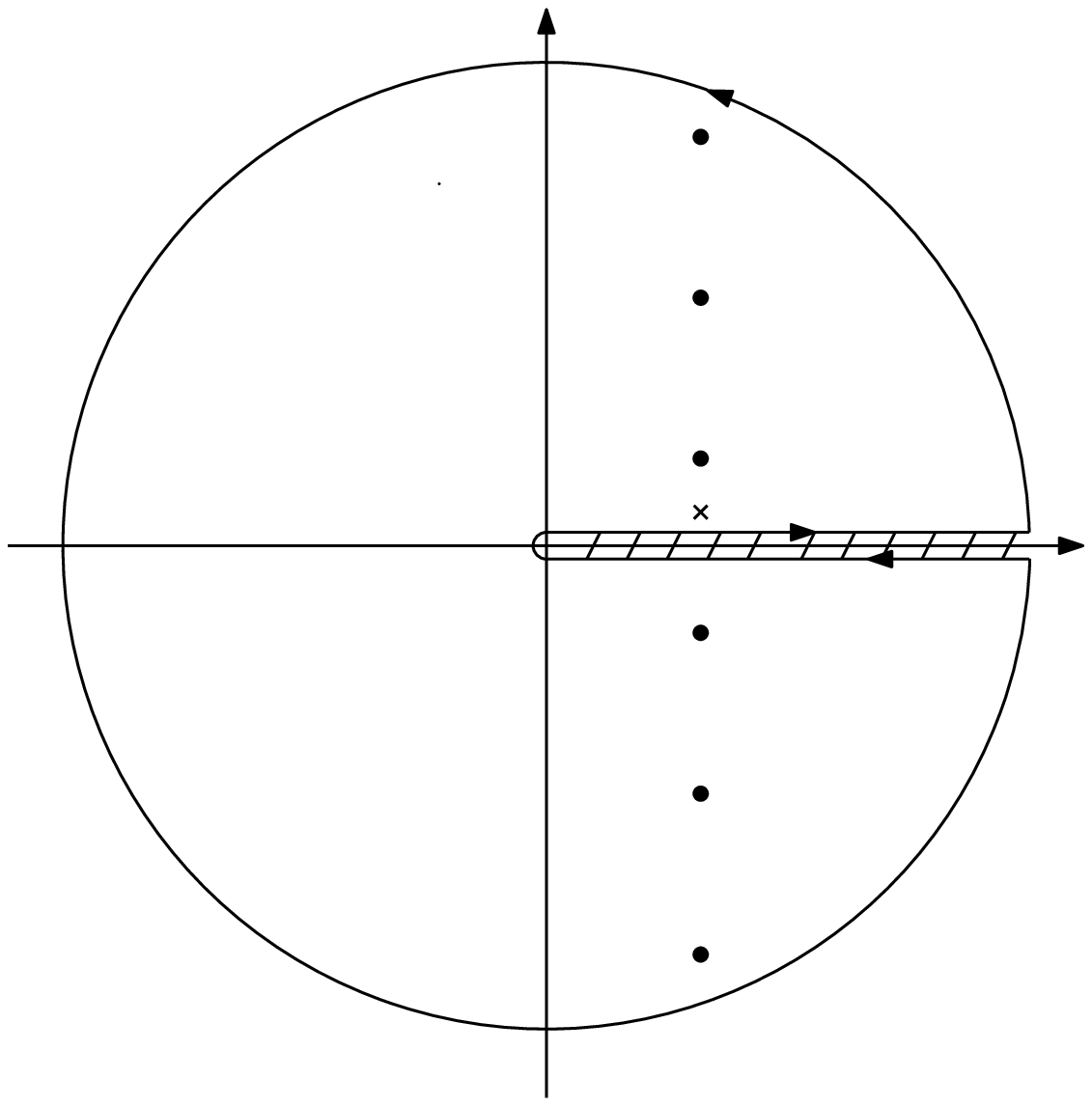}
  \caption{The contour ``c" in the calculation of Eq. (\ref{contour}). The dots ``$\cdot$" denote the first-order poles $z_n=\mu+\frac{(2n+1)\pi i}{\beta}$
  and the cross ``$\times$" denotes the second order pole $z_\eta=\mu+\eta i$. }
  \label{fig:contour}
  \end{center}
 \end{figure}
The contour integration can be evaluated in the summation of the
residuals as \bea &&\int_c
dz\sqrt{z}\frac{1-2N(z-\mu)}{(z-\mu-i\eta)^2}\cr=&&2\pi
i\Bigg[\lim_{z\rightarrow z_n}\frac{2\sqrt z/\beta }{(z-\mu-\eta
i)^2}+\lim_{z\rightarrow z_\eta}\frac{d}{dz}\Big[\sqrt
z\big(1-\frac{2}{\exp(\beta(z-\mu))+1}\big)\Big]
 \Bigg]\cr=&&2\pi i\Bigg[-\frac{2\beta\sqrt\mu}{\pi^2}\sum_{n=-\infty}^{+\infty}\frac{\sqrt{1+(2n+1)\pi i/\beta\mu}}{(2n+1)^2}+\sqrt{\mu}\frac{\beta}{2}\Bigg].\eea
Calculation shows that
$\sum_{n=-\infty}^{+\infty}\frac{\sqrt{1+(2n+1)\pi
i/\beta\mu}}{(2n+1)^2}$ is pure imaginary due to the symmetry of the
$z_n$ pole locations with respect to the horizontal axes. Then it
can be written as $\sum_{n=-\infty}^{+\infty}\frac{\sqrt{1+(2n+1)\pi
i/\beta\mu}}{(2n+1)^2}=\sqrt 2
i\sum_{n=0}^{+\infty}\frac{\sqrt{\sqrt{1+((2n+1)\pi/\beta\mu)^2}-\mbox{sgn}(\mu)}}{(2n+1)^2}.$
Hence, the parameter $u$ is calculated as \bea u
=\frac{(2m)^{3/2}}{16\pi^2}\Bigg[\frac{2\sqrt2\beta\sqrt{|\mu|}}{\pi}\sum_{n=0}^\infty\frac{\sqrt{\sqrt{1+(2n+1)^2(\frac{\pi}{\beta\mu})^2}-\mbox{sgn}(\mu)}}{(2n+1)^2}
-\frac{\pi\beta}{2}\sqrt{|\mu|}\theta(-\mu)+i\frac{\pi\beta}{2}\sqrt{|\mu|}\theta(\mu)\Bigg].\eea
In the same manner, the parameter $v$ can also be calculated as
shown in Eq. (6) and (7). We should note that while the expressions
for $u$ and others look different from the standard expression, as
given in ref.~\cite{Sademelosupp}, they in fact reduce to the same
expressions. We found that this form is more convenient to use the
above expression when dealing with higher order time-derivative
terms.

In the BCS and BEC limits all the parameters can be analytically
derived as shown in Table I.
\begin{table}[h]
\begin{center}
\begin{tabular}{|c|c|c|}\hline
\textbf{Parameters} & \textbf{BCS limit} & \textbf{BEC limit}\\
\hline $u'$ & 0 &
$\frac{\pi\nu(\epsilon_F)}{8\sqrt{\epsilon_F|\mu|}}$
\\
\hline $u''$ &$ \nu(\epsilon_F)\cdot\frac{\pi}{8k_BT}$&$0$
\\ \hline $v'$ & $\frac{7\nu(\epsilon_F)}{16\pi^2(k_BT)^2}\cdot\zeta(3)$ &
$\frac{\pi\nu(\epsilon_F)}{64\sqrt{\epsilon_F}|\mu|^{3/2}}$
\\
\hline $v''$ &$ -\nu(\epsilon_F)\cdot\frac{\pi}{32k_BT\epsilon_F}$&$0$ \\
\hline $\frac{1}{2m^\ast}$ &
$\frac{1}{2m}\cdot\frac{7\nu(\epsilon_F)\epsilon_F}{12\pi^2(k_BT)^2}\zeta(3)$
&
$\frac{1}{2m}\cdot\frac{\pi\nu(\epsilon_F)}{16\sqrt{\epsilon_F|\mu|}}$
\\ \hline $r$ & $\nu(\epsilon_F)\ln\frac{T_c}{T}$ & $\frac{\pi\nu(\epsilon_F)}{2\sqrt2\sqrt{\epsilon_F}}\cdot(\frac{1}{\sqrt m a_s}-\sqrt{2|\mu|})$\\\hline
$b$ & $\frac{7\nu(\epsilon_F)}{8\pi^2(k_BT)^2}\cdot\zeta(3)$
&$\frac{\pi\nu(\epsilon_F)}{32\sqrt{\epsilon_F}|\mu|^{3/2}}$\\
\hline
\end{tabular}
\end{center}
\caption{Asymptotic behaviors of the parameters in the
time-dependent Ginzburg-Landau theory in the BCS and BEC limits.}
\end{table}

\subsection{Spectral weight function in the case without damping
term} If we ignore the damping term by taking $u''=0$ the action can
be written as \be S=\int\frac{d\omega}{2\pi}\frac{d^3{\bf
k}}{(2\pi)^3}\bar{\Phi}(-\omega,-{\bf
k})\mathcal{G}^{-1}\Phi(\omega,{\bf k}), \ee with
$\bar{\Phi}(\omega,{\bf k})=(\delta_a(\omega,{\bf
k}),\delta_p(\omega,{\bf k}))$ and the kernel $\mathcal{G}$ is given
by
\begin{equation}
\mathcal{G}^{-1}=\left(
\begin{array}{cc}
-v'\omega^2+\xi_{\bf k}+2r & iu'\omega\\
-iu'\omega & -v'\omega^2+\xi_{\bf k}
\end{array}\right),
\end{equation}
with $k=|{\bf k}|$ and $\xi_{\bf k}=k^2/2m^*$. Then the
amplitude-amplitude correlation function can be easily calculated as
\bea \mathcal G_{aa}(\omega,{\bf k})=\frac{-v'\omega^2+\xi_{\bf
k}}{-u'^2\omega^2+(-v'\omega^2+\xi_{\bf k})(-v'\omega^2+\xi_{\bf
k}+2r)}. \eea Straight forward calculation yields the spectral
function as \bea A_{aa}(\omega,{\bf
k})&&=-\frac{1}{\pi}\mbox{Im}\mathcal G_{aa}(\omega+i\delta,{\bf
k})\cr&&=A_+( {\bf k})\delta(\omega-\omega_+({\bf k}))+A_-( {\bf
k})\delta(\omega-\omega_-({\bf k})),\eea where the mode frequencies
are given as \be \omega_\pm^2=\frac{\xi_{\bf
k}+r}{v'}+\frac{u'^2}{2v'^2}\pm\sqrt{\frac{r^2}{v'^2}+\frac{u'^4}{4v'^4}+\frac{u'^2}{v'^3}(\xi_{\bf
k}+r)} \ee and the spectra weight density \bea &&A_+( {\bf
k})=\frac{v'\omega_+^2-\xi_{\bf
k}}{2v'^2\omega_+(\omega_+^2-\omega_-^2)},\cr&& A_-( {\bf
k})=\frac{-v'\omega_-^2+\xi_{\bf
k}}{2v'^2\omega_-(\omega_+^2-\omega_-^2)}. \eea

At BCS limit the ratio of the two spectral weight densities can be
approximately calculated as \bea \frac{ A_-( {\bf k})}{A_+( {\bf
k})}=\frac{u'^2\sqrt{k^2/2m^\ast(k^2/2m^\ast+2r)}}{4v'r^2}.\eea At
BCS limit we have $u'/v'\rightarrow0$, this ratio vanishes. This
spectral weight transfer is shown in Fig. 2 in the main text.

\subsection{Spectral weight function in the case with damping term}
The spectral weight function of the amplitude mode in the case with
damping term $u''$ is \bea
A_{aa}&&=\frac{u''\omega}{2}\cdot\frac{|-v\omega^2+\frac{k^2}{2m^\ast}|^2+|u\omega|^2}{|-(u\omega)^2+(-v\omega^2+\frac{k^2}{2m^\ast})(-v\omega^2+\frac{k^2}{2m^\ast}+2r)|^2}\cr&&
=\frac{u''\omega}{2}\cdot\frac{|-v\omega^2+\frac{k^2}{2m^\ast}|^2+|u\omega|^2}{|v'^2(\omega^2-\tilde\omega_+^2)(\omega^2-\tilde\omega_-^2)-2iu'u''\omega^2|^2},\eea
where the eigen mode frequencies are
\bea\tilde\omega_\pm^2=\frac{\xi_{\bf
k}+r}{v'}+\frac{u'^2-u''^2}{2v'^2}\pm\sqrt{\frac{r^2}{v'^2}+\frac{(u'^2-u''^2)^2}{4v'^4}+\frac{u'^2-u''^2}{v'^3}(\xi_{\bf
k}+r)}.\eea For small momentum they can be approximated as \bea
&&\tilde\omega_-=\sqrt{\frac{2r\xi_{\bf
k}}{2v'r+u'^2-u''^2}},\cr&&\tilde\omega_+=\sqrt{\frac{2v'r+u'^2-u''^2}{v'^2}+\frac{2v'r+2u'^2-2u''^2}{v'(2v'r+u'^2-u''^2)}\xi_{\bf
k}}. \eea Compared with the case without damping term we see that
the gap of the Higgs mode is reduced from $\sqrt{2r/v'}$ to
$\sqrt{2r/v'-u''^2/v'^2}$ at BCS limit.

For small $\xi_{\bf k}$ the spectral weight on the Goldstone mode
can be calculated as \bea A_{aa}(\tilde\omega_-,{\bf
k})=\frac{u''}{8u'^2\tilde\omega_-}.\eea Different from the case
without damping term, we see that in the case with damping term the
spectral function has a weight proportional to $u''$ on the
Goldstone mode.

\subsection{The spectral weight function in the case with Coulomb
interaction} A time-dependent Ginzburg-Landau theory with Coulomb
interaction can be cast as~\cite{Schakelsupp} \bea F=\int dt d^3{\bf
x}\Big\{-\frac{1}{8\pi}\phi\nabla^2\phi+\bar\Delta\Big(-iu(\partial_t-2e\phi)+v(\partial_t-2e\phi)^2-\frac{\nabla^2}{2m^\ast}-r\Big)\Delta+\frac{b}{2}\bar\Delta
\bar\Delta\Delta\Delta\Big\},\eea where $e$ is the electric charge
and $\phi(t, {\bf x})$ is the electric field. By taking a symmetry
breaking $\Delta\rightarrow\Delta_0+\delta_a+i\delta_p$ we can have
a free energy for the low energy excitations in the momentum space
as \bea &&F=\int\frac{d\omega}{2\pi}\frac{d^3 {\bf
k}}{(2\pi)^3}\Bigg\{2ui\omega\delta_a(\omega,{\bf
k})\delta_p(-\omega,-{\bf k})+\delta_a(-\omega,-{\bf
k})(-v\omega^2+\frac{k^2}{2m^\ast}+2r)\delta_a(\omega,{\bf
k})+\delta_p(-\omega,-{\bf
k})\cr&&(-v\omega^2+\frac{k^2}{2m^\ast})\delta_p(\omega,{\bf
k})+4iue\Delta_0\phi(-\omega,-{\bf k})\delta_a(\omega,{\bf
k})-4ve\omega\Delta_0\phi(-\omega,-{\bf k})\delta_p(\omega,{\bf
k})+4ve^2\Delta_0^2\phi(-\omega,-{\bf k})\phi(\omega,{\bf
k})\Bigg\}.\cr&&\eea We integrate out the electric field $\phi$ and
obtain \bea &&F=\int\frac{d\omega}{2\pi}\frac{d^3 {\bf
k}}{(2\pi)^3}\Bigg\{2ui\omega\frac{k^2/8\pi}{k^2/8\pi+4ve^2\Delta_0^2}\delta_a(\omega,{\bf
k})\delta_p(-\omega,-{\bf k})+\delta_p(-\omega,-{\bf
k})(-v\omega^2\frac{k^2/8\pi}{k^2/8\pi+4ve^2\Delta_0^2}+\frac{k^2}{2m^\ast})\delta_p(\omega,{\bf
k})\Bigg\}\cr&&+\delta_a(-\omega,-{\bf
k})(-v\omega^2+\frac{k^2}{2m^\ast}+2r)\delta_a(\omega,{\bf k}).\eea
Then the spectral functions can be calculated as \bea &&{\rm
Im}\chi_{aa}=\frac{u''\omega}{2}\cdot\frac{(-v\omega^2\frac{k^2/8\pi}{k^2/8\pi+4ve^2\Delta_0^2}+\frac{k^2}{2m^\ast})^2+|u\omega\frac{k^2/8\pi}{k^2/8\pi+4ve^2\Delta_0^2}|^2}{|-(u\omega\frac{k^2/8\pi}{k^2/8\pi+4ve^2\Delta_0^2})^2+(-v\omega^2\frac{k^2/8\pi}{k^2/8\pi+4ve^2\Delta_0^2}+\frac{k^2}{2m^\ast})(-v\omega^2+\frac{k^2}{2m^\ast}+2r)|^2},\cr&&
{\rm
Im}\chi_{pp}=\frac{u''\omega}{2}\cdot\frac{(-v\omega^2+\frac{k^2}{2m^\ast}+2r)^2+|u\omega\frac{k^2/8\pi}{k^2/8\pi+4ve^2\Delta_0^2}|^2}{|-(u\omega\frac{k^2/8\pi}{k^2/8\pi+4ve^2\Delta_0^2})^2+(-v\omega^2\frac{k^2/8\pi}{k^2/8\pi+4ve^2\Delta_0^2}+\frac{k^2}{2m^\ast})(-v\omega^2+\frac{k^2}{2m^\ast}+2r)|^2}.\eea

\end{widetext}
\end{document}